\documentclass[manuscript,screen]{acmart}

\AtBeginDocument{%
  \providecommand\BibTeX{{%
    \normalfont B\kern-0.5em{\scshape i\kern-0.25em b}\kern-0.8em\TeX}}}

\usepackage{graphicx} 
\usepackage{stfloats} 
\usepackage{subfigure} 
\usepackage{enumitem}
\usepackage{multirow}
\usepackage{longtable}
\usepackage[most]{tcolorbox}
\usepackage{tabu}
\usepackage[ruled,linesnumbered]{algorithm2e} 
\usepackage{listings}
\usepackage{url}
\usepackage{amsmath}
\usepackage{amsfonts}
\usepackage{mathtools}
\usepackage{amsthm}
\usepackage{threeparttable}
\usepackage{caption}
\usepackage{color}
\usepackage{listings}
\usepackage{tcolorbox}

\definecolor{lightgreen}{rgb}{0.25, 0.63, 0.4375}
\definecolor{darkblue}{rgb}{0.02, 0.16, 0.49}

\definecolor{darkgreen}{rgb}{0, 0.5, 0}
\definecolor{darkred}{rgb}{0.72,0.04,0.04}

\setcopyright{acmcopyright}
\copyrightyear{2023}
\acmYear{2023}
\acmDOI{XXXXXXX.XXXXXXX}

\newcommand{\framework}{\textsc{WASMixer}}
\begin{document}

\title{{\framework}: Binary Obfuscation for WebAssembly}

\author{Shangtong Cao}
\authornote{Both authors contributed equally to this research.}
\email{shangtongcao@bupt.edu.cn}
\affiliation{%
  \institution{Beijing University of Posts and Telecommunications}
  \city{Beijing}
  \country{China}}

\author{Ningyu He}
\authornotemark[1]
\affiliation{%
  \institution{Key Lab on HCST (MOE), Peking University}
  \city{Beijing}
  \country{China}}
\email{ningyu.he@pku.edu.cn}

\author{Yao Guo}
\affiliation{%
  \institution{Key Lab on HCST (MOE), Peking University}
  \city{Beijing}
  \country{China}}
\email{yaoguo@pku.edu.cn}

\author{Haoyu Wang}
\affiliation{%
  \institution{Huazhong University of Science and Technology}
  \city{Wuhan}
  \country{China}}
\email{haoyuwang@hust.edu.cn}

\renewcommand{\shortauthors}{S. Cao and N. He, et al.}

\begin{abstract}
  WebAssembly (Wasm) is an emerging binary format that draws great attention from our community. However, Wasm binaries are weakly protected, as they can be read, edited, and manipulated by adversaries using either the officially provided readable text format (i.e., wat) or some advanced binary analysis tools. Reverse engineering of Wasm binaries is often used for nefarious intentions, e.g., identifying and exploiting both classic vulnerabilities and Wasm specific vulnerabilities exposed in the binaries. However, no Wasm-specific obfuscator is available in our community to secure the Wasm binaries. To fill the gap, in this paper, we present {\framework}, the first general-purpose Wasm binary obfuscator, enforcing data-level (string literals and function names) and code-level
  (control flow and instructions) obfuscation for Wasm binaries. We propose a series of key techniques to overcome challenges during Wasm binary rewriting, including an on-demand decryption method to minimize the impact brought by decrypting the data in memory area, and code splitting/reconstructing algorithms to handle structured control flow in Wasm. Extensive experiments demonstrate the correctness, effectiveness and efficiency of {\framework}. Our research has shed light on the promising direction of Wasm binary research, including Wasm code protection, Wasm binary diversification, and the attack-defense arm race of Wasm binaries.
\end{abstract}



\maketitle

\section{Introduction}
\label{sec:intro}
WebAssembly (Wasm) is a language-, platform-, and architecture-agnostic binary instruction format, proposed and endorsed by internet giants, including Google and Apple~\cite{bring-web-with-wasm}.
It aims to offer a both size- and load-time-efficient binary format, which
can execute at native speed while running on plenty of platforms, especially web browsers.
Therefore, its portability and efficiency make it a strong competitor for JavaScript. For example, lots of computation-intensive jobs are compiled to Wasm and run on browsers already, such as 3D graphic engines~\cite{3D}, cryptocurrency miners~\cite{deep-miner}, and multimedia encoders and decoders~\cite{media}. Furthermore, Wasm is moving fast towards a much wider spectrum of domains, e.g., mobile apps~\cite{mobile}, IoT~\cite{iot}, blockchain~\cite{eosio}, and serverless computing~\cite{Serverless}.

Though Wasm is a low-level binary format, it can be easily cracked by adversaries. 
On the one hand, to enable Wasm binaries to be read and edited by humans, Wasm officially provides a human-readable text format~\cite{call-indirect} (i.e., \texttt{wat}), which is an intermediate form designed to be exposed in text editors, browser developer tools, etc. 
On the other hand, reverse engineering of Wasm is quite easy by taking advantage of emerging binary analyzers for Wasm. For example, \textit{Manticore}~\cite{manticore} is a state-of-the-art static symbolic executor for Wasm. It is able to symbolically execute a function within a Wasm binary and explore multiple paths simultaneously in an abstract way. Through the final results, adversaries can obtain some original semantics of the binary, such as how it will respond against a specific set of inputs.
Furthermore, Wasm stores all string literals in a specific area in plain text, which may reflect the code semantics and developers' intentions.

Reverse engineering of Wasm binaries could be used for malicious intentions, e.g., to search for security vulnerabilities in Wasm binaries and based on which to launch attacks~\cite{crow}. In particular, a recent study~\cite{everything} suggests that many classic vulnerabilities are completely exposed in Wasm binaries, despite that they are no longer exploitable in native binaries due to common mitigations. 
Furthermore, many applications on other platforms can be ported to web browser with the help of Wasm, such as Photoshop~\cite{photoshop}, AutoCAD~\cite{autocad}, and some 3D games~\cite{games}. Code Obfuscation plays a crucial role in concealing the algorithms and essential logic employed by these software applications.
Considering the situation that Wasm binaries are weakly protected, code obfuscation for Wasm becomes an urgent need.

Though it is acceptable to conduct source-level obfuscation before compilation, it is impractical for the case of Wasm.
On the one hand, source-level obfuscation techniques are language-specific. They may vary in effectiveness after compiling to Wasm.
On the other hand, source-level obfuscation techniques cannot take full advantage of the characteristics of Wasm. For example, \textit{Tigress}~\cite{tigress}, a well-known obfuscator for C, only supports the Emscripten compiling chain~\cite{emscripten} instead of the Clang version~\cite{wasi-sdk}, which is often used to compile standalone Wasm binaries. Moreover, it cannot obfuscate the names of imported and exported functions, due to its ineffectiveness on Emscripten's libraries. 
For this reason, it is necessary to implement a binary-level obfuscator against Wasm.

However, implementing a Wasm-specific obfuscator is challenging, which can be summarized in three-fold.
First, obfuscations should bring in as little as overhead in terms of executing time and binary size due to the high efficiency and compact format of Wasm. For example, inserting a function to decrypt all encrypted string literals before executing or at runtime should consider both these two factors.
Second, Wasm adopts a highly complicated structured control flow. There are no \texttt{goto}-like instructions in Wasm, and such a structure only allows control flow directed in one-way. Implementing control flow obfuscations, such as flattening, while keeping the consistency of semantics is challenging.
Third, there are several Wasm-specific static analysis frameworks proposed recently, some of which, especially the symbolic executors, can recover code semantics from obfuscated Wasm binaries. Even introducing opaque predicates, which are effective to evade human inspection, is not enough to bypass those symbolic executors.

\textbf{This Work.} 
In this paper, to the best of our knowledge, we present the first Wasm-specific obfuscation framework, named {\framework}.
It is composed of two main modules, i.e., \textit{data obfuscator} and \textit{code obfuscator}.
In general, the data obfuscator is responsible for replacing all function names with random strings, and decrypting all pre-encrypted string literals in Wasm binaries on demand at runtime, which brings in less overhead than one-time decryption at loading. As a complementary, the code obfuscator manipulates instructions and the control flow to resist both human reverse engineering and static analysis, like alias disruption, control flow flattening, and Collatz-based opaque predicates that can resist static analysis.
Based on benchmarks consisting of representative Wasm binaries, the evaluation results prove the effectiveness of {\framework}.
Specifically, it suggests that obfuscating methods in {\framework} will not bring in any negative impact on the syntactic correctness and semantic consistency. Moreover, {\framework} can also effectively resist manual reverse engineering, hide original intents, and hinder state-of-the-art static analyzers while introducing less than 20\% overhead in both terms of executing time and binary size.

\textbf{Our contribution} can be summarized as follows:
\begin{itemize}
	\item \textbf{The first general-purpose Wasm binary obfuscator.} To the best of our knowledge, we have proposed the first language-agnostic binary obfuscation solution for Wasm, named {\framework}, which consists of two modules, specifically against data (string literals and function names) and code (control flow and instructions), respectively. 
	\item \textbf{Key techniques for Wasm binary rewriting.} We have proposed a series of key techniques to deal with Wasm binaries, including an on-demand decryption method to minimize the impact brought by decrypting the data in memory area, and code block splitting algorithm and reconstructing interfaces to handle structured control flow in Wasm.
	\item \textbf{Extensive evaluation and useful application scenarios.} We have performed extensive experiments to evaluate the correctness, effectiveness and efficiency of {\framework}, suggesting that it can evade human analysis, state-of-the-art Wasm analyzers and commercial anti-virus engines. 
\end{itemize}

To boost further research on Wasm binaries, we will release {\framework}, along with all the benchmark datasets, to the community.

\section{Background}
\label{sec:background}
In this section, we will briefly illustrate some basic concepts of Wasm as well as code obfuscation.

\subsection{WebAssembly (Wasm)}
\label{sec:backgroud:wasm}
Wasm is an emerging language that can be regarded as the compilation target of multiple mainstream programming languages, e.g., C/C++~\cite{c/c++}, Rust~\cite{Rust}, and Go~\cite{Go}.
Its advantages mainly reside in its native-like execution speed and compact binary size.
We next briefly introduce its features relevant to this paper.

\noindent
\textbf{Types \& Instructions.}
There are only four primitive value types defined in Wasm, i.e., \texttt{i32}, \texttt{i64}, \texttt{f32}, and \texttt{f64}.
The \textit{i} and \textit{f} refer to \textit{integer} and \textit{float}, respectively, while the number corresponds to the length in bits.
Wasm has defined over 170 instructions~\cite{instrs} which can consume operands from and produce a return value onto its \textit{operand stack}.
For example, \texttt{i32.const} pushes its immediate number, expressed as a 32-bit integer, onto the stack.

\noindent
\textbf{Data Structure.}
Wasm has designed a set of simple but effective data structures.
Specifically, all data structures in Wasm adopt \textit{key-value} mechanism.
For example, each function has its owned \textit{local} structure, local values can be accessed by \texttt{local.get i}, where \texttt{i} is the index.
Data in \textit{global} and \textit{memory} can be accessed and shared by all functions. However, only four primitive types can be stored in the \textit{global} structure, while non-scalar types, e.g., string and array, are stored in the \textit{memory} area.
Wasm adopts linear memory, i.e., a string of continuous bytes.
A set of instructions are used to load and store data with the memory. For example, \texttt{i32.load addr} will take four continuous bytes as a 32-bit integer starting from \texttt{addr}.

\noindent
\textbf{Wasm Binary.}
A Wasm binary is composed of \textit{sections}, each of which has its specific functionality.
Moreover, a section is regarded as a vector of elements.
For example, the \textit{import section} consists of a set of elements, and each element corresponds to an imported function with its name and index.
Wasm officially offers a set of tools~\cite{wasm2wat, wat2wasm} to support lossless translation between the binary format and a \textit{WebAssembly text format} (wat)\footnote{In this paper, all code snippets are wat files converted from Wasm binaries.}.

\noindent
\textbf{Control Flow.}
Wasm supports both \textit{direct call} and \textit{indirect call} by \texttt{call} and \texttt{call\_indirect}, respectively.
To be specific, a \texttt{call} will consume necessary arguments from the stack according to its designated callee, and push the return value (if it has) onto the stack.
Instead of directly designating the callee, \texttt{call\_indirect} takes the top element on the stack as the function index of the callee.

\begin{figure}[t]
\centering
\includegraphics[width=0.5\columnwidth]{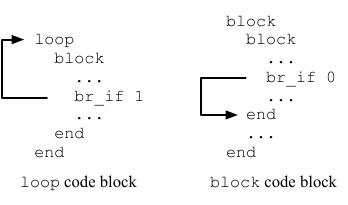}
\caption{Two valid control flow jumping in Wasm.}
\label{fig:cfg}
\end{figure}

Additionally, Wasm applies a specific \textit{structured control flow}.
Specifically, instructions in Wasm are divided into \textit{code blocks}, which can be led by a \texttt{block} or \texttt{loop} instruction.
Code blocks can be nested, but the context of an inner code block is independent to its outer one.
Moreover, there are no \texttt{goto}-like instructions in Wasm, indicating the control flow cannot be directed to arbitrary instructions.
Fig.~\ref{fig:cfg} illustrates the only two allowed control flow jumping rules.
As we can see, \texttt{br\_if} is a conditional jump, i.e., the jump will be performed only if the top element on the stack is not zero. The number following the jump instruction refers to how many layers are intended to jump out (0 is the current one, 1 is its parent, and so on).
If the destination code block is led by a \texttt{loop}, the control flow will be directed to its heading, or, led by a \texttt{block}, the control flow can only go to the tailing \texttt{end}.

\subsection{Obfuscation}
\label{sec:backgroud:obfuscation}
Obfuscation is a widely used software protection method that textually translates programs while preserving semantics to protect the program from being cracked or to prevent sensitive data leakage~\cite{obfuscation}.
Obfuscation methods can be roughly divided into two categories: \textit{data obfuscation} and \textit{code obfuscation}.
Specifically, the data obfuscation mainly focuses on readable literals (e.g., function names and string literals) in programs, and translates them into semantic-equivalent but unreadable (or meaningless) ones~\cite{data-obfuscation}.
The code obfuscation often plays on instructions and the control flow. For example, replacing a single arithmetic instruction with a sequence of bit-shifting instructions, or constructing bogus control flows within a critical function~\cite{fake-control}.

Obfuscation takes not only humans as its counterparty but also program analysis techniques.
Currently, except for the readability, whether an obfuscated program can resist automated program analysis techniques (e.g., symbolic execution) also becomes one of the criteria for evaluating the effectiveness of the obfuscation method~\cite{schwartz2010all}.
There are generally two kinds of principles~\cite{two-principle}.
One is to deliberately introduce path explosion issue by constructing complicated control flow. To this end, symbolic executors have to place significant time and resources on meaningless paths constructed by the obfuscator.
The second is to exploit the backend SMT solver, like introducing non-linear formula, to make it hard for symbolic executors to determine whether a path is feasible or not.

\section{{\framework}}
\label{sec:overview}
In this section, we first overview the architecture of {\framework}, and then illustrate the necessity of implementing such a binary-level obfuscator.
Last, we also detail the challenges in designing and implementing {\framework}.

\begin{figure*}[t] 
\centering 
\includegraphics[width=0.9\textwidth]{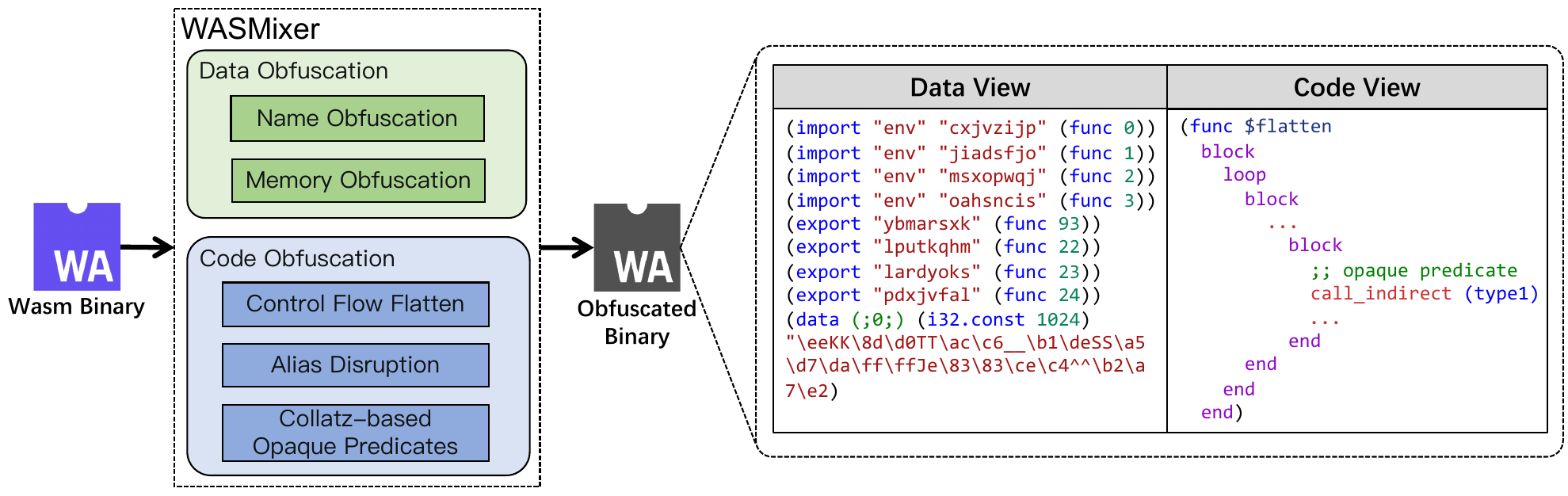}
\caption{The architecture and workflow of {\framework}.}
\label{fig:chaos}
\end{figure*}

\subsection{Overview}
{\framework} takes a Wasm binary as input and generates an obfuscated binary based on the specified options, as shown in Fig.~\ref{fig:chaos}.
As we can see, {\framework} is mainly composed of two components: \textit{data obfuscator} and \textit{code obfuscator}.
Data obfuscator can perform name and memory obfuscation. The purpose of name obfuscation is to replace names of functions, import and export items, and other identifiers with random strings, while memory obfuscation encrypts the initialized linear memory in Wasm binary and decrypts it at runtime.
Moreover, {\framework} also proposes three code obfuscation methods, including flattening obfuscation for control flow, alias disruption for \texttt{call} instructions, and Collatz-based opaque predicates for resisting static analysis, e.g., symbolic execution. These obfuscation methods can effectively resist both human reverse engineering and program analysis.
The implementation of {\framework} is detailed in \S\ref{sec:approach}.

\subsection{Why Binary Obfuscator?}
\label{sec:overview:compare}

\def\baddef/{\textcolor{darkred}{\bf $\times$}}
\def\verybaddef/{\textcolor{darkred}{\bf $\times\times$}}
\def\gooddef/{\textcolor{darkgreen}{{\bf \checkmark}}}

Compared with source code and compiler IR level obfuscators, the biggest advantage of obfuscating Wasm binaries is the \textit{compatibility}.
Specifically, on the one hand, source code level obfuscation inherently has limitations when source code is unavailable, especially when dealing with a large number of legacy binaries. In addition, this lack of direct connection to the source code isolates the source semantics, making it more challenging for attackers to analyze. For instance, the relationship between Wasm instructions can be converted to an Abstract Syntax Tree (AST) through stack operations, where the ASTs hold source code semantics. Binary obfuscation can construct uncommon stack instruction snippets that obfuscate the semantics of the generated AST.
On the other hand, obfuscating methods for diverse source programming languages are different from each other, resulting in significant differences upon obfuscating effectiveness on compiled Wasm binaries. In general, obfuscators for compiled languages have more methods to protect programs, while compression and encryption are more commonly used obfuscation methods for scripted languages. Binary obfuscation is not limited by language type and can provide obfuscations at the same level across all languages.
Moreover, source code level obfuscation inherently has limitations when source code is unavailable, especially when dealing with a large number of legacy binaries. 
Moreover, as mentioned in \S\ref{sec:approach:code-obfuscation:alias}, the Wasm language has its own unique and effective binary obfuscation methods that are not covered by previous obfuscators.

\begin{table}[t]
\caption{Comparison between Tigress and {\framework}.}
    \begin{tabular}{lcc}
    \toprule
    \textbf{Functions}                 & \textbf{Tigress} & \textbf{{\framework}} \\ \midrule
    support multi-languages    &  \baddef/    & \gooddef/  \\ 
    unavailable source code             &  \baddef/    & \gooddef/  \\ 
    supported compilers        &  \textcolor{darkred}{emscripten}  & \textcolor{darkgreen}{all}  \\ 
    obfuscating import \& export &  \baddef/    & \gooddef/  \\ 
    obfuscating debug information &  \baddef/    & \gooddef/  \\ 
    against symbolic execution &  \baddef/    & \gooddef/  \\ \bottomrule
    \end{tabular}
    \label{table:compare}
\end{table}

Take one of the most popular obfuscation tools for C programming language, \textit{Tigress}~\cite{tigress}, as an example.
For clarification, we summarize the pros and cons of Tigress and {\framework} in Table~\ref{table:compare}.
We can see that in addition to compatibility issues, i.e., multi-language supporting and unavailability of source code, we also find that Tigress can only support the Emscripten~\cite{emscripten} compiler, without supporting compilers like clang~\cite{wasi-sdk}, which is often used for compiling standalone Wasm binaries.
Additionally, Tigress is unable to obfuscate the library provided by Emscripten, which indicates that both the import and export items and the debug information cannot be replaced by meaningless strings.
Last, although Tigress provides complex obfuscation methods such as virtualization, they can still not resist symbolic execution, like KLEE~\cite{klee}.
Consequently, our binary-level obfuscator can fill the gap and overcome the inherent shortcomings of source code level obfuscators.

\subsection{Challenges}
\label{sec:overview:challenges}
To the best of our knowledge, we have implemented the first binary obfuscator against Wasm binaries. Although there are several existing efforts~\cite{wasabi,wasm-mutate,SEISMIC,fuzzm} on Wasm binary analysis, they are limited to instruction-level rewriting, e.g., Wasabi~\cite{wasabi} and Wasm-mutate~\cite{wasm-mutate}, or simple control flow translation, e.g., Fuzzm~\cite{fuzzm}.
Especially, in implementing {\framework}, we face mainly three challenges.

\noindent
\textbf{C1: Low-overhead \& General Data Obfuscation.}
Except for the implementation and declaration of functions, a Wasm binary contains lots of data, e.g., initial memory data and debugging information.
Effectively obfuscating these data without affecting semantic consistency is challenging.
For instance, inserting a decryption function between the initiation and the execution of a Wasm binary will significantly increase the response time, which is harmful to users' experience.
Or, the decryption can be performed on demand during executing \texttt{load} instructions, e.g., by substituting or hooking these instructions. However, there are more than 20 \texttt{load}- and \texttt{store}-related instructions in Wasm for different bit-length. Implementing and inserting multiple hooking functions in a single Wasm binary is impractical and inefficient.
Additionally, Wasm does not mandate the format of various types (e.g., variable types and function names) of debugging information in its \textit{custom section}, which may vary according to different source languages.

\noindent
\textbf{Our Solution:}
To minimize the impact brought by decrypting the data in memory area, we propose an on-demand decryption method.
For different memory loading and storing instructions, the method can dynamically obtain the destination address, and determine how many bits should be loaded or stored. Then, a runtime decryption or encryption will be conducted accordingly.
As for obfuscating the function names in debugging information, we also design a parsing method that can precisely extract function names with variable length, and replace them with random characters without breaking the original structure.

\noindent
\textbf{C2: Structured Control Flow.}
As we illustrated in \S\ref{sec:backgroud:wasm}, Wasm adopts the structured control flow, where the control flow can only be jumped sequentially, i.e., from an inner code block to outer ones.
Moreover, destinations of jump instructions cannot be designated arbitrarily. They have to be the head (led by \texttt{loop}) or the tail (led by \texttt{block}) of the targeted code block.
Without the help of \texttt{goto}-like instructions, it is challenging to modify the control flow in Wasm to perform obfuscation.
Moreover, code blocks are independent to each other. In other words, variables cannot be shared across code blocks. To this end, some obfuscating methods that require splitting code blocks, e.g., control flow flattening~\cite{flatten}, cannot be conducted easily while keeping the semantics intact.

\noindent
\textbf{Our Solution:}
An important prerequisite for accomplishing code obfuscation is to ensure semantic consistency.
With this mandatory requirement in mind, we first propose a code block splitting algorithm, which can split a given code block into several ones ensuring the stack balance of new blocks and variable sharing across blocks meanwhile.
Then, we propose another algorithm, named code block rearranging algorithm, which can rearrange the split code blocks in a flattened manner.
To this end, the control flow of any given Wasm code blocks can be flattened, which achieves the goal of obfuscation.

\noindent
\textbf{C3: Effective Obfuscation against Static Analysis.}
There are two counterparties for program obfuscation, i.e., human reverse engineering and static analyzers.
Increasing unreadability, e.g., memory encryption or function names obfuscation, on Wasm binaries may be effective for countering human, but not for static analyzers, e.g., symbolic executor.
It is because symbolic executors can recover some semantic information to some extent. For example, the result of an opaque predicate cannot be easily obtained by a hacker, but it is a piece of cake for symbolic executor.
Therefore, we should deliberately invalidate symbolic execution on Wasm level.

\noindent
\textbf{Our Solution:}
In order to solve this problem, we decided to take advantage of the inherent limitation of symbolic execution, i.e., path explosion.
We will introduce path explosion by two methods.
On the one hand, we replace all static calls with indirect calls, which has to force symbolic executors to try all possible callees because the callee will only be determined at runtime.
On the other hand, we introduce an opaque predicate, in which it has an unbounded loop. And we deliberately pass an undetermined value (symbol) as the input of this opaque predicate. To this end, symbolic executors have to maintain an exponentially growing number of paths at every calling point to this opaque predicate.

\section{Approach} 
\label{sec:approach}
In this section, we will depict the technical details of data obfuscator and code obfuscator in {\framework}.

\subsection{Data Obfuscator}
\label{sec:approach:data-obfuscation}
As we mentioned in \S\ref{sec:backgroud:obfuscation}, data obfuscation is used to increase the difficulty of reading and understanding the data by either human or automated analyzers.
For Wasm binary, the data can be roughly divided into the linear memory and the debugging information.

\subsubsection{Memory Obfuscation}
\label{sec:approach:data-obfuscation:memory}
In a Wasm binary, memory data is stored in plaintext. It mainly composed of constant strings in the source program, like string templates in \texttt{printf} and \texttt{scanf}~\cite{memory}.
To some extent, such a way of storing data in plaintext will inevitably reveal the code semantics and developers' intentions, facilitating the cracking of Wasm binaries.
It is critical, therefore, to obfuscate the memory data declared initially or even generated dynamically.

To decrypt the encrypted memory data, two strategies can be applied: \textit{decrypting at the entry} or \textit{decrypting on-demand}.
Specifically, as for the former method, a decryption function will be inserted as a wrapper at the entry function. Once the binary is initiated, its memory data will be decrypted all at once. However, it will introduce a huge overhead.
On the one hand, not all data will be retrieved during an actual execution. Such decrypting all data will consume unnecessary time and resources.
On the other hand, decrypting with the binary initiation will increase its response time, decreasing the overall experience.
Note that, the data generated during the runtime is not encrypted. Therefore, if the runtime memory can be read through certain vulnerabilities (e.g., out-of-bound reading~\cite{everything}), it can also lead to data leakage.

\begin{algorithm}[t]
    \caption{On-demand \& Runtime Memory Loading and Storing Algorithm.}
    \label{algo:load-store}
    \SetAlgoLined
    \DontPrintSemicolon
    \SetKwFunction{load}{dec\_load}\SetKwFunction{store}{enc\_store}
    \SetKwProg{myproc}{Procedure}{}{end}
    \KwIn{\textit{base} - base address of target, \\
    \textit{offset} - offset address of target, \\
    \textit{signed} - padding as signed or not, \\
    \textit{len} - the length of loaded data in bits, \\
    \textit{type} - the target type}
    \KwOut{\textit{data} - loaded data}
    \myproc{\load{base, offset, signed, len, type}}{
    $address \gets base + offset$\;
    $data \gets \texttt{loadMemory}(address, len)$\;
    $key \gets \texttt{getKey}()$\;
    $data \gets \texttt{xor}(data, key)$\hfill\{Decryption\}\\
    \eIf{$signed$}
    {
        $data \gets \texttt{signedExtend}(data, type)$\;
    }{
        $data \gets \texttt{unsignedExtend}(data, type)$\;
    }
    \Return{$data$}\;
    }
    \vspace{0.05in}
    \myproc{\store{base, offset, signed, type}}{
    $key \gets \texttt{getKey}()$\;
    $data \gets \texttt{xor}(data, key)$\hfill\{Encryption\}\\
    $data \gets \texttt{trunc}(data, type)$\;
    $address \gets base + offset$\;
    $\texttt{storeMemory}(address, type, data)$\;
    }
\end{algorithm}

To resolve the \textbf{C1} (see \S\ref{sec:overview:challenges}), we propose an algorithm that can load (store) data from (to) memory on-demand with a runtime encryption to guarantee the data confidentiality, as shown in Algorithm~\ref{algo:load-store}.
As it shows, we first replace all \texttt{load} and \texttt{store} instructions as \texttt{call dec\_load} and \texttt{call enc\_store}, respectively.
Let us take the \texttt{dec\_load} as an example.
There are 14 \texttt{load} instructions defined in the Wasm specification. The instruction \texttt{i32.load8\_u} will load 1 byte from the target address, extend it as a 4-byte unsigned integer, and push it onto the stack.
Therefore, in the implementation of \texttt{dec\_load}, we first calculate the target address by \texttt{base} and \texttt{offset}, both of which are pushed onto the stack already.
Then, after loading the data according to \texttt{len} (L3), the key will be loaded and XOR-ed with the loaded data (L5). Because the XOR operation is symmetric, the key is generated randomly and kept in both \texttt{dec\_load} and \texttt{enc\_store}.
At last, the loaded data will be extended according to the target type.
The process plays similarly in \texttt{enc\_store}.
Note that, at L15, we also conduct the XOR operation on the key with the to-be-stored data, which keeps the confidentiality of runtime data.

\subsubsection{Name Obfuscation} 
\label{sec:approach:data-obfuscation:name}
Name obfuscation aims to rename identifiers, e.g., variables and functions, to avoid malicious cracking and prevent data leakage~\cite{nameOB}.
In Wasm, there are two types of identifiers should be protected, i.e., readable function names in debugging information, and function identifiers in import/export sections.

Debugging information in Wasm is stored in the custom section, which is generated by compilers directly.
It consists of some sensitive metadata of the binary, e.g., compiler version, function names, variable types, and even the path of its source code~\cite{debug-info}.
The format of function names is specified as follows:
$$\texttt{\textbf{name}}||\texttt{\textbf{0x1}}||len_{sec}||\overbrace{num_{name}||\underbrace{(idx_i||len_{name_i}||name_i)}_\text{repeat $num_{name}$ times}}^\text{length in $len_{sec}$ bytes}$$
, where the leading \texttt{name} and \texttt{0x1} are string literals, and the length of each function name is declared by its corresponding $len_{name_i}$ \cite{name-section}.
Therefore, to hinder reverse engineering, we keep everything intact except for $name_i$.
We replace them with random strings with the same length.
In this way, execution logic and call relationships for all functions are identical to the original ones, but function names have been changed to meaningless strings for human.

Wasm allows the import functions from or exports its implementation of functions to its environment.
These two parts are declared in the import and export section, respectively.
Take the export section as an example.
For each declaration in the export section, it is composed of $name_{internal}$ and $name_{export}$, i.e., exporting the internal function, dubbed $name_{internal}$, as $name_{export}$.
To this end, the outer environment (like web browsers) can call the function $name_{export}$ through JavaScript~\cite{wasm-link}.
However, the plaintext of $name_{export}$ can somewhat reflect its intention, which may be used as tokens by detectors~\cite{memory}.
We offer an option to substitute random characters for $name_{export}$.
It is worth noting that a small side effect of this obfuscating method is that it requires modifying outer modules to call the random strings instead of $name_{export}$.
Thus, we offer an auxiliary tool to assist developers to update the call to $name_{export}$ as the random string in its outer environment.

\subsection{Code Obfuscator}
\label{sec:approach:code-obfuscation}
As we mentioned in \S\ref{sec:backgroud:obfuscation}, code obfuscation plays on instructions and control flow.
In this section, we will introduce how we deal with Wasm-specific instructions and its structured control flow.

\subsubsection{Control Flow Obfuscation}
\label{sec:approach:code-obfuscation:flattening}
Control flow obfuscation~\cite{control-flow-ob} is a set of mainstream and effective obfuscating methods that complicate program's control flow to make it unreadable and hard to analyze. Typically, it involves inserting new control flows to confuse analyzers, or modifying and complicating existing control flows.
Due to the complicated structured control flow adopted by Wasm, it is challenging to achieve that.
Hence, we abstract performing control flow obfuscation in two major steps, i.e., \textit{code block splitting} and \textit{code block rearranging}.
Except for detailing these two steps, we also demonstrate how to take advantage of them to conduct \textit{control flow flattening}~\cite{flatten}.

\paragraph{Stage I: Code Block Splitting.}
The structured control flow applied by Wasm makes it challenging for performing code block splitting as we mentioned in \textbf{C2}.
Specifically, in Wasm, code blocks are independent to each other. In other words, variables on stack that can be accessed by a code block cannot be retrieved after splitting it into several ones.
In summary, splitting instructions into several code blocks requires support specifically against Wasm.

\begin{algorithm}[t]
    \caption{Code Block Splitting Algorithm.}
    \label{algo:code-block-split}
    \SetAlgoLined
    \DontPrintSemicolon
    \KwIn{\textit{cb} - the target code block, \
    \textit{num} - the number of code blocks intended to split}
    \KwOut{\textit{cbs} - generated code blocks}
    $funcIdx \gets \texttt{extractFuncId}(cb)$\;
    $maxLen \gets \texttt{getMaxStackLength}(cb)$\;
    $\texttt{appendFuncLocal}(funcId, maxLen)$\;
    $cbs \gets \texttt{listDivide}(cb, num)$\;
    \ForEach{$cb$ in $cbs$}{
        $stackPost \gets \texttt{getPostStack}(cb)$\;
        $stackPre \gets \texttt{getPreStack}(cb)$\;
        \ForEach{$type$ in $stackPost$}{
            $\texttt{modifyLocalType}(funcId, type)$\;
            $\texttt{insertInstruction}(cb, type, \texttt{local.set})$\hfill\{Store stack\}\\
        }
        \ForEach{$type$ in $stackPre$}{
            $\texttt{modifyLocalType}(funcId, type)$\;
            $\texttt{insertInstruction}(cb, type, \texttt{local.get})$\hfill\{Restore stack\}\\
        }
    }
    \Return{$cbs$}\;
\end{algorithm}

\begin{figure}[t]
\centering  
\includegraphics[width=0.5\columnwidth]{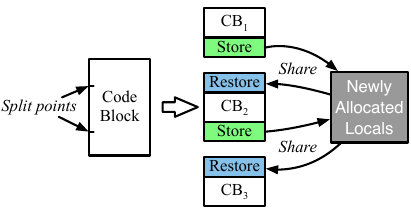}
\caption{Split a code block into separate ones, among which the arguments are shared through newly allocated locals.} 
\label{fig:split-code-block} 
\end{figure}

\begin{figure*}[t] 
    \flushleft  
    \includegraphics[width=\textwidth]{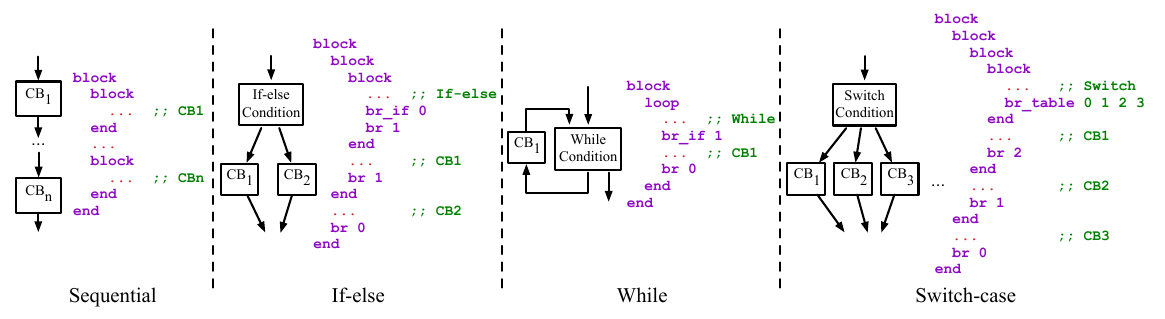} 
    \caption{Four basic relationships among code blocks.} 
    \label{fig:control-flow-api} 
\end{figure*}

We propose a code block splitting algorithm, which is shown in Algorithm~\ref{algo:code-block-split}.
It takes a code block and a number that refers to how many blocks are intended to split as inputs.
The algorithm firstly extracts which function the target block locates in (L1), and calculates the maximum size of used stack slots according to its contained instructions (L2). To this end, it can determine how many locals (as temporary variable sharing area between newly split code blocks) should be newly allocated in the function (L3).
After splitting the code block into several ones, storing and restoring stack elements should be conducted after and before each of them, respectively.
As illustrated in Fig~\ref{fig:split-code-block}, each newly split code block is ended with a list of instructions to store necessary variables, and started with instructions that restore them.
Take the stack storing as an example, the algorithm will traverse a newly generated code block to determine the number and the types of elements remained, which are maintained in $stackPost$ at L6.
According to the types of remained elements, L9 modifies the type of the corresponding local.
An extra \texttt{local.set} will be appended to the code block, thus remained elements can be temporarily stored in local variables, which can be accessed by the restoring process of the next code block.
Consequently, a given code block is split into several tiny code blocks, among which variables could be shared between newly allocated locals.

\paragraph{Stage II: Code Block Rearranging.}
The code block splitting algorithm can only split a code block into several ones, where they still keep a sequential relationship.
As we mentioned in \textbf{C2}, there is no \texttt{goto}-like instructions that can arbitrarily guide the control flow. The instructions \texttt{br} and \texttt{br\_if} can only jump to designated code blocks' beginning or tailing (see \S\ref{sec:backgroud:wasm}).
It is challenging and requisite to implement a set of interfaces that allow users to assemble given code blocks according to various relationships on demand without considering the correctness of syntax.
To this end, we have summarized four basic relationships among code blocks, i.e., \textit{sequential}, \textit{if-else}, \textit{while}, and \textit{switch-case}, as shown in Fig.~\ref{fig:control-flow-api}.

As we can see, the basic elements to compose these four relationships are code blocks and a condition (except for the sequential one).
In other words, once given enough elements, a structure following the designated relationship will be assembled.
For example, the interface of if-else relationship is:
$$\texttt{assembleIfElse(cond, cb1, cb2)}$$
, where \texttt{cond} is a condition statement, and \texttt{cb1} and \texttt{cb2} are code blocks either constructed by users or generated by the code block splitting algorithm.
In either case, users can complete the rearrangement without concerning details of syntactic correctness.

\paragraph{Case Study: Control Flow Flattening.}
Based on the code block splitting algorithm and interfaces exposed by code block rearranging, we can easily implement the control flow flattening obfuscation.
In general, control flow flattening is to split a series of instructions into code blocks, which are cases for a switch-case statement. Such a switch-case will be wrapped by a loop to execute recursively till all cases are executed once in the original order.
Thus, we have implemented a flattening obfuscation, which is shown in Algorithm~\ref{algo:flatten}.
Suppose it takes a series of split code blocks as input.
Specifically, we first record the original order of the given code blocks and shuffle them for enhancing the obfuscation effect (L1 and L2).
To pass the index of the to be executed code block, we have to allocate another local variable, which is done by L3 and L4.
Then, for each code block, we append a list of instructions (after the variable storing, see Fig.~\ref{fig:split-code-block}) that can jump back to the switch condition, and pass an index that refers to the successive block through an unconditional jump (L5 to L10).
Afterward, we construct a switch condition (achieved by \texttt{br\_table}), and assemble it with the code blocks mentioned above into a switch-case statement (L11 and L12).
Moreover, we should further wrap the switch-case into a while statement to achieve iterating on each case (L13 and L14).
To this end, each code block will be executed once in the original order.

\begin{algorithm}[t]
    \caption{Control Flow Flattening.}
    \label{algo:flatten}
    \SetAlgoLined
    \DontPrintSemicolon
    \KwIn{\textit{cbs} - a list of code blocks}
    \KwOut{\textit{flattenCB} - a code block composed of given code blocks with flattened relationship}
    $cbsOrder \gets \texttt{recordOrder}(cbs)$\;
    $\texttt{shuffle}(cbs)$\;
    $funcIdx \gets \texttt{extractFuncId}(cbs)$\;
    $jumpFlagIdx \gets \texttt{appendFuncLocal}(funcIdx)$\;
    \ForEach{$cb$ in $cbs$}{
        $postCB \gets \texttt{findSuccCB}(cbsOrder, cb)$\;
        $\texttt{appendLocalAssign}(cb, jumpFlagIdx, postCB)$\;
        $nestedLevel \gets \texttt{getNestedLevel}(cbsOrder, cb)$\;
        $\texttt{appendJumpInstr}(cb, nestedLevel)$\;
    }
    $switchCond \gets \texttt{createSwitchCond}(jumpFlagIdx, cbsOrder)$\;
    $switchCaseCB \gets \texttt{assembleSwitchCase}(switchCond, cbs)$\;
    $whileCond \gets \texttt{appendExit}(switchCaseCB, jumpFlagIdx)$\;
    $flattenCB \gets \texttt{assembleWhile}(whileCond)$\;
    \Return{$flattenCB$}
\end{algorithm}

\begin{figure}[t] 
\center 
\includegraphics[width=0.5\columnwidth]{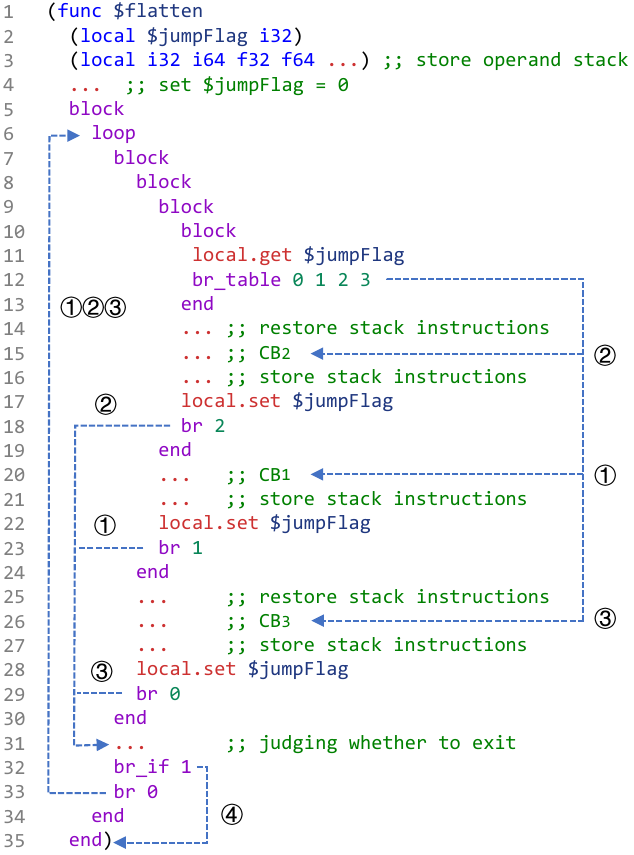}
\caption{A code snippet to illustrate a flattened control flow.} 
\label{fig:flatten-eg} 
\end{figure}

Fig.~\ref{fig:flatten-eg} shows a concrete example of the structure of flattened code blocks.
As we can see, three code blocks are given and flattened in an order different from the original one.
For each block, it begins with variable restoring instructions, and ends with variable storing instructions as well as instructions that designate its successive code block.
In this example, the original sequential order is shuffled, i.e., the first case (L14 to L18) will be executed at the second place, which increases the difficulty for human reverse engineering to some extent.
Note that, according to the \textit{while} structure defined in Fig.~\ref{fig:control-flow-api}, both the switch-case statement and the exit instruction (L31) are the condition of the while statement, while its body is an empty code block, which should be inserted between L32 and L33.
Our rearranging interfaces regard such empty code blocks as valid situations.
Consequently, we can conclude that by combining the code block splitting algorithm and the interfaces of code block rearranging, we can achieve code block flattening in Wasm.

\subsubsection{Alias Disruption Obfuscation} 
\label{sec:approach:code-obfuscation:alias}
As we mentioned in \S\ref{sec:backgroud:wasm}, in Wasm, function invocation can be achieved using either \texttt{call} or \texttt{call\_indirect}, while the former one, static call, is widely used and can be easily analyzed.
Replacing the former with the latter one can achieve alias disruption obfuscation to some extent.

Specifically, a static call will designate its callee directly, like \texttt{call \$foo}. 
However, the indirect call, named \texttt{call\_indirect}, calls a function according to the top element of the stack at runtime.
It is designed for function pointers and polymorphism in high-level languages. As explained in Daniel~\cite{indirectcall}, \texttt{call\_indirect} is hard to analyze for determining the index value of an indirect function is challenging. Since the type of index value is \texttt{I32}, which has no unique characteristic to assist analysis.
Furthermore, such a runtime-determined callee will also hinder the static analysis, especially for enumerating possible callees~\cite{call-indirect}.
In a Wasm binary, there is a special section, named \textit{elem section}, where it declares all possible callees of \texttt{call\_indirect} by a list of function references at a specific offset. 
To perform inter-procedural analysis, the static analyzer has to enumerate all possible callees declared in the \textit{elem section}.

To achieve such an alias disruption obfuscation, many steps are required.
First, we need to obtain the indices of callees of all \texttt{call} instructions, and append them into the \textit{elem section}.
Then, before each to-be-replaced \texttt{call} instructions, we insert an \texttt{i32.const} instruction with its callee's index.
Finally, we will examine the type of the callee, and pass the type index as the argument of the replaced \texttt{call\_indirect}\footnote{Pass the callee's type index is mandatory for \texttt{call\_indirect} to guarantee the stack balance.}.
For example, a \texttt{call \$foo} can be replaced by:
\begin{align*}
	&\texttt{i32.const 7} \\
	&\texttt{call\_indirect (type 2)}
\end{align*}
, where we assume the function \texttt{\$foo} is the 7th function, and its type is declared as the 2nd one.

To further improve the obfuscating effect, we take advantage of the opaque predicate \cite{opaque-predicate} instead of an \texttt{i32.const} instruction.
Specifically, an opaque predicate refers to an expression whose result is constant regardless of the input. Typically, the returned value of an opaque predicate is only known to its obfuscator, and is difficult for analyzers to infer. A simple opaque predicate, e.g., \texttt{x\*(x-1)\%2}, will always return 0 regardless of the value of \texttt{x}.
We can replace some of \texttt{i32.const} before \texttt{call\_indirect} instructions into opaque predicates to improve the effectiveness of the alias disruption obfuscation.

\subsubsection{Collatz-based Opaque Predicate Obfuscation}
\label{sec:approach:code-obfuscation:collatz}
Currently, in Wasm, several static symbolic execution tools have been proposed, like manticore~\cite{manticore} and WANA~\cite{WANA}. Theoretically, with the help of symbolic executors, attackers can retain the semantics of Wasm binaries easily, which may invalidate the obfuscation.
To this end, we have implemented a collatz-based opaque predicate obfuscating method in Wasm to evade the analysis of such tools.

One of the mainstream methods to counter symbolic execution is deliberately introducing path explosion.
For example, introducing an unbounded loop that has no effect on original semantics.
We introduced \textit{Collatz conjecture}~\cite{collatz-conjecture}, a famous unsolved problem in mathematics also known as 3n + 1 conjecture, to achieve this goal.
The conjecture asserts whether repeating two simple arithmetic operations will eventually transform every positive integer into 1.
Specifically, the two operations are:
$$collatz(n)=\left\{\begin{array}{ll}
n / 2 & \text { if } n \equiv 0 \\
3 n+1 & \text { if } n \equiv 1
\end{array} \quad(\bmod\ 2)\right.$$
, where the conjecture takes a positive integer and checks whether it is odd or even.
Most importantly, there is no general term formula for Collatz conjecture. In other words, we cannot determine how many rounds are required for the final convergence on 1.

Note that, although our fellow researchers have applied the Collatz conjecture to code obfuscation in previous work~\cite{collatz}, we are the first to implement it in Wasm.
During symbolically executing the function $collatz$, if the input is or contains any symbolic values, it will conduct path forking on whether the given $n$ is odd or even.
After several rounds, the number of paths will increase exponentially, which requires lots of computing- and space-resources.
Therefore, we use it as an opaque predicate to construct the \textit{jumpFlag} in both control flow flattening (see \S\ref{sec:approach:code-obfuscation:flattening}) and alias disruption obfuscation (see \S\ref{sec:approach:code-obfuscation:alias}).
Take constructing \texttt{jumpFlag} in Fig.~\ref{fig:flatten-eg} as an example, and we assume the value of \texttt{jumpFlag} to be assigned is 3.
To this end, the assignment expression for the \texttt{jumpFlag} is:

\lstset{
    numbers=left,
    numberstyle=\tiny,
    basicstyle=\ttfamily\small,
}

\begin{lstlisting}[]
a = m * x + c
while (collatz(a) > 1):
	a = collatz(a)
jumpFlag = collatz(a) + 2
\end{lstlisting}
, where $x$ is a symbol, which generally comes from one of the parameters of the function, and $m$ and $c$ are all random integers.
To this end, the final value of \texttt{jumpFlag} will always be 3 regardless of the values of $x$, $m$ and $c$.
However, in this way, the function \texttt{collatz} will take a monomial as input, which can be easily solved if the context of symbolic variable is already constrained. For example, if the constraint of $x$ is already limited in $\left[0,3\right]$, only four paths need to be searched.
In order to increase the search space and thus reduce the analysis efficiency, we can replace the $a$ at L1 with a polynomial, like:
\begin{align*}
	a = m*x + n*y + c
\end{align*}
, where $y$ is also a parameter in the function. Even if $y$ is limited in a constraint, e.g., $\left[5,8\right]$, the number of paths is the Cartesian product of possibilities of $x$ and $y$. Thus, there are 16 (4*4) paths are required to be searched.
If it is necessary, more terms can be inserted into the polynomial. 
Consequently, symbolically executing this piece of code is extremely inefficient due to the exponentially increased number of paths imported by the Collatz conjecture.

\section{Implementation \& Evaluation}
\label{sec:imple-eval}

\subsection{Implementation}
We have implemented {\framework} with over 5.9K LOC of Python3 code from scratch.
To avoid reinventing the wheel, we implement the Wasm binary decoder based on wasm-python-book\footnote{\url{https://github.com/Relph1119/wasm-python-book} (commit: 0x872bc8f)}, a simple but effective Wasm runtime.
We have packaged {\framework} into a standard Python library. Based on the exposed obfuscating methods, security developers can conveniently apply them to specified Wasm binaries.

\subsection{Research Questions \& Experimental Setup}
\label{sec:evaluation:rq}
Our evaluation is driven by the following three research questions:

\begin{itemize}
    \item[\textbf{RQ1}] Can those implemented obfuscating methods in {\framework} keep semantic consistency?
    \item[\textbf{RQ2}] How effective are the provided obfuscation methods against human reverse engineering, malware detectors and static analyzers?
    \item[\textbf{RQ3}] How much overhead are brought by {\framework} in terms of runtime and code size?
\end{itemize}

To answer the above RQs, we have collected a sophisticated set of Wasm binaries, which are shown in Table~\ref{table:benchmark}.
Specifically, to answer \textbf{RQ1}, we need some binaries that have explicit and determined input and output. Basic Algorithms~\cite{basic-algorithm} (\textbf{D$_1$}) is composed of some commonly used algorithms, like string concatenation and quick sort on an array, compiled from C to Wasm. Moreover, we also collected two real-world applications (base64 and md5) found on wapm~\cite{wapm} (\textbf{D$_2$}), a well-known and mainstream Wasm package manager.

To answer \textbf{RQ2}, we have collected three datasets (\textbf{D$_3$} to \textbf{D$_5$}) and three representative analyzers against Wasm binaries (VirusTotal, manticore, and wasp).
To be specific, we chose VirusTotal~\cite{VirusTotal}, the most authoritative and widely-used malware analysis service currently, which is supported by more than 60 security vendors, i.e., each target will be scanned by more than 60 detectors. Based on SEISMIC~\cite{SEISMIC} and MinerRay~\cite{minerray}, we have collected 18 Wasm binaries that are labeled as malicious cryptominers as the ground truth, which dataset is denoted as \textbf{D$_3$}.
As for the static analyzers, we chose two state-of-the-art symbolic executors, i.e., \textit{manticore}~\cite{manticore} and \textit{wasp}~\cite{wasp}.
Specifically, manticore is a commercial and actively maintained native symbolic executor for Wasm binaries. It provides several features, e.g., fine-grained control of state exploration via event callbacks.
Different from manticore, wasp is a concolic executor. In other words, it dynamically executes the given Wasm binary with random inputs and records all path conditions. According to the solutions returned by its backend SMT-solver, wasp can mutate the seed and reach a high coverage.
To evaluate the effectiveness of our proposed obfuscating methods against symbolic execution, we enforced both manticore and wasp to execute cases from the btree-traverse~\cite{btree} (\textbf{D$_5$}). These cases perform a process of initiating, inserting, traversing, and deleting on btrees with different orders. Moreover, Gillian-Collections-C~\cite{gillian} (\textbf{D$_4$}) is a test suite for testing the Collections-C~\cite{gillian-c}, a well-known library for common data structures implemented in C. However, because wasp only supports a subset of Wasm instructions, only manticore is set to analyze \textbf{D$_4$}.
Last, we have evaluated its ability to resist human reverse engineering on all \textbf{D$_3$}, \textbf{D$_4$} and \textbf{D$_5$}, using a number of metrics.

Last, to answer \textbf{RQ3}, we measured the imported overhead on all datasets except for \textbf{D$_3$}. This is because these 18 cryptominers all run on browsers, requiring lots of dependency modules and interactions with the corresponding JavaScripts.
Therefore, we only measure the overhead on those standalone cases.

\begin{table}[t]
    \caption{Datasets used by different RQs.}
    \resizebox{0.6\columnwidth}{!}{%
        \begin{tabular}{ccc}
            \toprule
                                                                                       & \textbf{Datasets}                                              & \textbf{\#Wasm Binaries} \\ \midrule
            \multirow{2}{*}{RQ1}                                                       & Basic Algorithms~\cite{basic-algorithm} (\textbf{D$_1$})                      & 18                       \\
                                                                                       & wapm programs~\cite{wapm} (\textbf{D$_2$})                         & 2                        \\	\midrule
            \begin{tabular}[c]{@{}c@{}}RQ2\\ (VirusTotal)\end{tabular}                 & cryptomining programs~\cite{SEISMIC,minerray} (\textbf{D$_3$})                 & 18                       \\	\midrule
            \multirow{2}{*}{\begin{tabular}[c]{@{}c@{}}RQ2\\ (manticore)\end{tabular}} & Gillian-Collecitons-C~\cite{gillian} (\textbf{D$_4$})                 & 159                      \\
                                                                                       & btree-traverse~\cite{btree} (\textbf{D$_5$})                        & 32                       \\	\midrule
            \begin{tabular}[c]{@{}c@{}}RQ2\\ (wasp)\end{tabular}                       & btree-traverse~\cite{btree} (\textbf{D$_5$})                        & 32                       \\	\midrule
            RQ3                                                                        & \textbf{D$_1$}, \textbf{D$_2$}, \textbf{D$_4$}, \textbf{D$_5$} & 243                      \\ \bottomrule
        \end{tabular}%
    }
    \label{table:benchmark}
\end{table}

All experiments were performed on a server running Ubuntu 22.04 with a 64-core AMD EPYC 7713 CPU and 256GB RAM.

\subsection{RQ1: Semantic Consistency}
\label{sec:evaluation:rq1}
As an obfuscator, maintaining semantic consistency is a necessary condition that must be met. To this end, we must evaluate if a Wasm binary can keep identical semantics before and after the obfuscation of {\framework}.
In total, we choose 20 cases from \textbf{D$_1$} and \textbf{D$_2$} (see Table~\ref{table:benchmark}) as candidates.
Specifically, all these 20 cases can generate explicit and determined output when given a set of inputs. For each of them, we generate 14 different obfuscated versions according to different obfuscating options. The two options are applying memory and name obfuscation (see \S\ref{sec:approach:data-obfuscation}), respectively, while the other 12 options are divided into two groups, the only difference between these two groups is enabling the Collatz-based opaque predicate obfuscation (see \S\ref{sec:approach:code-obfuscation:collatz}) or not.
Further, each group consists of 6 cases, corresponding to the number of split code blocks as 5, 10, and 20 in control flow flattening obfuscation (see \S\ref{sec:approach:code-obfuscation:flattening}) and performing alias disruption obfuscation (see \S\ref{sec:approach:code-obfuscation:alias}) on 25\%, 50\%, and 100\% candidate instructions.

For each candidate case, along with its 14 mutated versions, we construct sets of inputs to cover as many program paths as possible. For example, for an argument type as \texttt{unsigned char}, we will randomly choose its value ranging from 0 to 255. For a string, we will construct a string with random printable characters whose length is from 0 to 10.
Moreover, as each Wasm binary will be statically verified before executing~\cite{static-validate}, we also asked wasm-validate\footnote{An official tool to verify the validity of Wasm binaries.} to examine the syntactic correctness of these 300 cases.

Results show that with different obfuscating options, all 300 obfuscated Wasm binaries can still pass the verification of wasm-validate, which indicates that {\framework} will not corrupt the syntax of given Wasm binaries.
Moreover, with identical sets of input, obfuscated cases can generate the same output as the original one. We can conclude that {\framework} will not bring in any negative impact towards syntax and semantics during obfuscating.

\begin{tcolorbox}[title= \textbf{RQ-1} Answer, left=2pt, right=2pt, top=2pt, bottom=2pt]
    By combining and applying 5 proposed obfuscating methods on 20 Wasm binaries with determined output, results show that {\framework} will not bring in any negative impact on the correctness of syntax and semantic consistency.
\end{tcolorbox}

\subsection{RQ2: Effectiveness}
The effectiveness of obfuscation can be evaluated from three perspectives.
First, obfuscated Wasm binaries should be unreadable for malicious users who typically can arbitrarily access them.
Second, obfuscated Wasm binaries can effectively hide their original intents.
Third, obfuscated Wasm binaries should be hard for analyzing by static analyzers. For example, for a static symbolic executor, the analyzing time for an obfuscated Wasm binary should be unacceptable.
We evaluate the effectiveness of {\framework} in the following.

\begin{figure}[!t]
    \centering  
    \includegraphics[width=0.7\columnwidth]{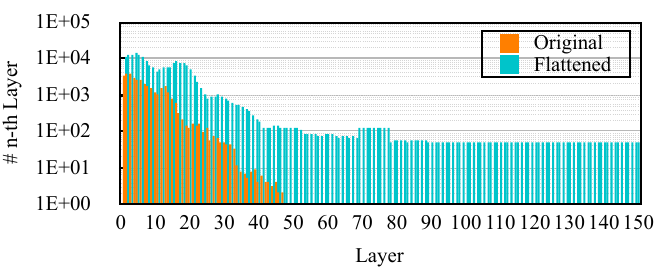}
    \caption{The distribution of $n$-th nesting layers before and after control flow flattening for Wasm binaries in \textbf{D$_1$} to \textbf{D$_5$}.}
    \label{fig:layer_count} 
\end{figure}

\begin{table}[]
    \centering
    \caption{The number of \texttt{call\_indirect} and the length of element section for Wasm binaries in \textbf{D$_4$} and \textbf{D$_5$} before and after \texttt{call} to \texttt{call\_indirect} obfuscation.}
    \label{table:callindirectcount}
    \resizebox{0.6\columnwidth}{!}{%
        \begin{tabular}{@{}ccccc@{}}
            \toprule
            \textbf{}                                     & \multicolumn{2}{c}{\textbf{\#\texttt{call\_indirect}}} & \multicolumn{2}{c}{\textbf{length of the elem section}}                                                                           \\
                                                          & Original                                    & \begin{tabular}[c]{@{}c@{}}Obfuscated (times)\end{tabular} & Original & \begin{tabular}[c]{@{}c@{}}Obfuscated (times)\end{tabular}
            \\ \midrule
            \begin{tabular}[c]{@{}c@{}}\textbf{D$_1$}\end{tabular}    & 234                                        & 3,180 (13.6x)                                                      & 72     & 1,110 (15.4x)                                                      \\
            \begin{tabular}[c]{@{}c@{}}\textbf{D$_2$}\end{tabular}    & 188                                        & 5,423 (28.8x)                                                      & 253     & 963 (3.8x)                                                      \\
            \begin{tabular}[c]{@{}c@{}}\textbf{D$_3$}\end{tabular}    & 92                                        & 2,153 (23.4x)                                                      & 282     & 1,446 (5.1x)                                                      \\
            \begin{tabular}[c]{@{}c@{}}\textbf{D$_4$}\end{tabular}    & 3,554                                        & 32,998 (9.3x)                                                      & 1,159     & 11,868 (10.2x)                                                      \\
            \begin{tabular}[c]{@{}c@{}}\textbf{D$_5$}\end{tabular}    & 0                                        & 606 (N/A)                                                     & 0     & 198 (N/A)                                                     \\
            \begin{tabular}[c]{@{}c@{}}\textbf{Total}\end{tabular} & \textbf{4,069}                                        & \textbf{44,360 (10.9x)}                                                      & \textbf{1,766}     & \textbf{15,585 (8.8x)}                                                      \\
            \bottomrule
        \end{tabular}%
    }
\end{table}

\subsubsection{Against Manual Reverse Engineering}
\label{sec:evaluation:rq2:manual}
As we mentioned in \S\ref{sec:backgroud:wasm}, for a valid Wasm binary, users can convert it to a readable text file, i.e., wat file, through an official toolkit.
Four obfuscating methods can significantly increase the unreadability of a wat file, i.e., \textit{name obfuscation}, \textit{memory obfuscation}, \textit{control flow flattening}, and \textit{alias disruption}.
Because it is obvious that name and memory obfuscation can resist manual reverse engineering effectively, we only evaluate the other two methods in this section.
Specifically, the flattening will deliberately split a code block and insert lots of jump instructions. Such a frequent and nested control flow jumping between code blocks will definitely confuse malicious users.
Moreover, by applying alias disruption, once we convert \texttt{call} to \texttt{call\_indirect}, its operand, i.e., the callee, will not be statically designated. In other words, malicious users have to consider all functions listed in the elem section as possible callees (see \S\ref{sec:approach:code-obfuscation:alias}).

Therefore, we have applied both methods on all cases from \textbf{D$_1$} to \textbf{D$_5$}.
Figure~\ref{fig:layer_count} illustrates the distribution of \textit{n}-th nesting layers after the control flow flattening is conducted.
As we can see, there is an explicit shift towards the right after performing flattening, which indicates that the number of layers with high orders increases.
The results show that, for \textit{n}-th nesting layers ($n > 5$), the numbers for cases in \textbf{D$_1$} to \textbf{D$_5$} increase by 7.14 times.
Moreover, before flattening, the highest $n$ for is 72 (occurs in a case in {D$_3$}), while it turns to 489 after flattening.
It can be proven that \textit{control flow flattening can significantly increase the number of high-order nesting code blocks, which makes Wasm binaries unreadable and hard to follow}.

We also measure the number of \texttt{call\_indirect} after performing alias disruption obfuscation, which is shown in Table~\ref{table:callindirectcount}.
We can easily observe that the number of \texttt{call\_indirect} increases by 10.9 times after the obfuscation, while the length of the corresponding elem section increases by 8.8 times.
Interestingly, there is no \texttt{call\_indirect} instructions in cases in \textbf{D$_5$}, because all its belonging cases are manually constructed~\cite{wasp}. After converting all their \texttt{call} instructions, we have got cases with more than 600 \texttt{call\_indirect} instructions with identical semantics.
It can be concluded that \textit{alias disruption obfuscating method can effectively convert \texttt{call} to \texttt{call\_indirect}. And the length of elem section increases by at least an order of magnitude, which can significantly increase the workload for malicious users because they have to consider all of them as possible callees for each \texttt{call\_indirect} instruction}.

\subsubsection{Against Malware Detector}
\label{sec:rq2:malware}

Due to the superior performance of Wasm over JavaScript~\cite{bring-web-with-wasm}, many malicious users secretly inject cryptominers written in Wasm into the victim webpages, or deploy a miner on its owned pages.
Once a user unintentionally visits these pages, cryptominers will be executed by utilizing the computing resources (e.g., CPU and hard disks) of the visitors to gain profits for the deployers.
However, current malware detectors, e.g., VirusTotal, can effectively and efficiently identify these cryptominers' intentional malicious intents, and report them as malware~\cite{VirusTotal}.
One of the responsibilities of obfuscation is to hide the original intents of the given programs. To this end, we applied all obfuscating methods except for the Collatz-based one to 18 cases in \textbf{D$_3$}. Table~\ref{table:VT} shows the results of how many security vendors in VirusTotal can identify these malware.

\begin{table}[t]
    \centering
    \caption{The results of VirusTotal on identifying cryptomining Wasm binaries with different obfuscation options. OB$_1$ to OB$_5$ refer to name obfuscation, memory obfuscation, control flow flattening, alias disruption, and the combination of name and memory obfuscation, respectively.}
    \label{table:VT}
    \resizebox{0.6\columnwidth}{!}{%
        \begin{tabular}{@{}cclllll@{}}
            \toprule
            \multirow{2}{*}\textbf{} & \textbf{} & \multicolumn{3}{c}{\textbf{Name Obfuscation}} & \multicolumn{2}{c}{\textbf{Code Obfuscation}} \\ \cmidrule(l){3-7}
                \textbf{ID} & \textbf{original} & \multicolumn{1}{c}{\textbf{OB$_1$}} & \multicolumn{1}{c}{\textbf{OB$_2$}} & \multicolumn{1}{c}{\textbf{OB$_5$}} & \multicolumn{1}{c}{\textbf{OB$_3$}} & \multicolumn{1}{c}{\textbf{OB$_4$}}
            \\ \midrule
            dd8aabc0    & 32                & 15 (53\%)            & 7  (78\%)             & 0 (100\%)         &  32 (41\%)    & 19 (0\%)               \\ 
            8e2fc626    & 19                & 2  (89\%)            & 11 (42\%)             & 0 (100\%)         &  14 (26\%)    & 14 (26\%)              \\ 
            eeeaaf2e    & 17                & 3  (82\%)            & 11 (35\%)             & 0 (100\%)         &  13 (12\%)    & 15 (24\%)              \\ 
            74a4ed58    & 14                & 3  (79\%)            & 2  (86\%)             & 0 (100\%)         &  7  (50\%)    & 7  (50\%)              \\ 
            9cda640f    & 7                 & 0  (100\%)           & 1  (86\%)             & 0 (100\%)         &  4  (14\%)    & 6  (43\%)              \\ 
            b170c14e    & 7                 & 2  (71\%)            & 3  (57\%)             & 0 (100\%)         &  5  (14\%)    & 6  (29\%)              \\ 
            a2b93b31    & 6                 & 2  (67\%)            & 0  (100\%)             & 0 (100\%)        &  5  (17\%)    & 5  (17\%)              \\ 
            73c6c519    & 5                 & 2  (60\%)            & 0  (100\%)             & 0 (100\%)        &  3  (0\%)     & 5  (40\%)             \\ 
            50ca16c9    & 4                 & 2  (50\%)            & 1  (75\%)             & 0 (100\%)         &  4  (0\%)     & 4  (0\%)              \\ 
            e742774b    & 3                 & 2  (33\%)            & 0  (100\%)             & 0 (100\%)        &  2  (33\%)    & 2  (33\%)              \\ 
            37ff6a30    & 3                 & 0  (100\%)           & 0  (100\%)             & 0 (100\%)        &  3  (0\%)     & 3  (0\%)              \\ 
            41087519    & 2                 & 2  (0\%)             & 0  (100\%)             & 0 (100\%)        &  2  (0\%)     & 2  (0\%)              \\ 
            1f925c46    & 2                 & 2  (0\%)             & 0  (100\%)             & 0 (100\%)        &  2  (0\%)     & 2  (0\%)              \\ 
            5bec2bbb    & 2                 & 2  (0\%)             & 0  (100\%)             & 0 (100\%)        &  2  (0\%)     & 2  (0\%)              \\ 
            f879f0ce    & 2                 & 2  (0\%)             & 0  (100\%)             & 0 (100\%)        &  2  (0\%)     & 2  (0\%)              \\ 
            64709eac    & 2                 & 2  (0\%)             & 0  (100\%)             & 0 (100\%)        &  2  (0\%)     & 2  (0\%)              \\ 
            c0d72dbb    & 2                 & 2  (0\%)             & 0  (100\%)             & 0 (100\%)        &  1  (50\%)    & 1  (50\%)              \\ 
            d8c98da8    & 2                 & 2  (0\%)             & 0  (100\%)             & 0 (100\%)        &  2  (0\%)     & 2  (0\%)              \\ 
            \bottomrule
        \end{tabular} %
    }
\end{table}

As we can see, without any obfuscating methods, every cryptominer can be identified by at least two security vendors, and the one with id \texttt{dd8aabc0} can even be detected by 32 detectors. 
Interestingly, we can observe that code obfuscations (\textbf{OB$_3$} and \textbf{OB$_4$}) has little contribution in hiding original intents, especially the control flow flattening. None of these 18 cases can escape VirusTotal.
Contrarily, name obfuscating methods perform very well.
By simply applying name obfuscation and memory obfuscation independently, 2 and 11 cases have escaped all detectors in VirusTotal, respectively, accounting for 11.1\% and 61.1\% of all cases.
Moreover, if we combine these two methods, all 18 cases cannot be identified by any detectors at all.
To this end, we can conclude that most of detectors in VirusTotal identify malicious Wasm binaries by matching function names and string literals, instead of recovering semantics by static analysis.

\subsubsection{Against Static Analysis}
\label{sec:evaluation:rq2:static}
As we mentioned in \S\ref{sec:evaluation:rq}, we choose manticore and wasp as representatives for static analyzers. To this end, the effectiveness of {\framework} against static analysis is equivalent to whether an obfuscated Wasm binary can be analyzed by symbolic executors within an acceptable range in terms of both consumed time and resources.
In \S\ref{sec:approach:code-obfuscation:collatz}, we have introduced the Collatz-based opaque predicates that can be integrated into the control flow flattening and alias disruption. Both of these can deliberately introduce path explosion due to the characteristics of Collatz conjecture.
Table~\ref{table:effectiveness} illustrates the consumed time on obfuscated Wasm binaries of cases in \textbf{D$_4$} and \textbf{D$_5$}.

\begin{table}[]
    \centering
    \caption{Consumed time on analyzing Wasm binaries of \textbf{D$_4$} and \textbf{D$_5$} with different obfuscating options for manticore and wasp, where $O_1$ and $O_2$ indicate the integration of Collatz-based opaque predicates, and TO refers to timeout (48 hours).}
    \label{table:effectiveness}
    \begin{tabular}{@{}cccc@{}}
        \toprule
        \multirow{2}{*}{\textbf{}}                                                                    & \multicolumn{2}{c}{\textbf{manticore}} & \textbf{wasp}                   \\ \cmidrule(l){2-4}
                                                                                                      & \textbf{D$_4$}                         & \textbf{D$_5$} & \textbf{D$_5$} \\ \midrule
        original                                                                                      & 0.28h                                  & 12.3h          & 0.5h           \\ \midrule
        flattening                                                                                    & 0.28h                                  & 12.3h          & 0.5h           \\
        \begin{tabular}[c]{@{}c@{}}flattening ($O_{1}$)\end{tabular}       & TO (0/159)                             & TO (0/32)      & TO (0/32)      \\
        \begin{tabular}[c]{@{}c@{}}flattening ($O_{2}$)\end{tabular}       & TO (0/159)                             & TO (0/32)      & TO (0/32)      \\ \midrule
        alias disruption                                                                              & 0.28h                                  & 12.3h          & 0.5h           \\
        \begin{tabular}[c]{@{}c@{}}alias disruption ($O_{1}$)\end{tabular} & TO (0/159)                             & TO (0/32)      & 0.6h           \\
        \begin{tabular}[c]{@{}c@{}}alias disruption ($O_{2}$)\end{tabular} & TO (0/159)                             & TO (0/32)      & 0.75h          \\ \bottomrule
    \end{tabular}
\end{table}

As we can see, the first column is the obfuscating methods adopted, where $O_2$ indicates a more aggressive obfuscation than $O_1$. For example, for the flattening, $O_1$ only inserts two Collatz-based opaque predicates for each function, while $O_{2}$ divides every ten instructions into a code block and appends a predicate.
The following columns refer to consumed time, where timeout (TO) indicates the dataset cannot be finished within 48 hours.
Comparing the 3rd, 4th and 7th row, we can see that adopting simple control flow flattening and alias disruption without integrating Collatz-based opaque predicates can only improve the unreadability (see \S\ref{sec:evaluation:rq2:manual}) rather than the resistance against symbolic execution.
However, even if adopting a less aggressive option ($O_1$), both manticore and wasp cannot solve even one of the cases from either \textbf{D$_4$} or \textbf{D$_5$} within two days at all.
This is because once a function is called, the two Collatz predicates should be solved. However, the number of paths grows exponentially due to its inherent characteristic.
Interestingly, for wasp, both the $O_1$ and $O_2$ Collatz-based alias disruption can only extend the consumed time by 20\% and 50\%, respectively.
We think the reason behind such an ineffectiveness is twofold.
First, the wasp is a concolic symbol executor, which is inherently more efficient than static symbol executors like manticore.
Second, the cases of \textbf{D$_5$} are not directly compiled from source code written in high-level programing language, but generated by artificial construction, which significantly decreases the number of \texttt{call} instructions (as shown in Table~\ref{table:callindirectcount}). To this end, it has fewer instrumented points for inserting the Collatz opaque predicates.
We have to argue that the manually constructed Wasm binaries are extremely unusual in real-world environments.
Based on the results, we can conclude that \textit{both the Collatz-based control flow flattening and alias disruption can effectively hinder the performance of symbolic executors}.

\begin{tcolorbox}[title= \textbf{RQ-2} Answer, left=2pt, right=2pt, top=2pt, bottom=2pt]
    After adopting different obfuscation methods on cases from \textbf{D$_1$} to \textbf{D$_5$}, results show that {\framework} can provide sufficient capabilities to resist manual reverse engineering, hide original intents, and hinder state-of-the-art static analyzers.
\end{tcolorbox}

\subsection{RQ3: Overhead}
\label{sec:rq3}
Except for standalone Wasm binaries, they are often uploaded and used in web browsers as a library, e.g., graphic computation~\cite{wasm-webgpu} and cryptocurrency mining~\cite{mining}.
Therefore, users will be sensitive to their size and runtime performance, which are related to loading consumed time and executing time, respectively.
As side effects of obfuscating, introducing overheads in terms of binary size and runtime are inevitable. But we should minimize their effects as small as possible.
To evaluate them, we adopt all mentioned obfuscating options on all datasets except for \textbf{D$_3$}, because of its strong dependency on environment (see \S\ref{sec:evaluation:rq}).
Table~\ref{table:overhead} illustrates the measured results.

\begin{table}[]
    \centering
    \caption{Overheads of consumed time and binary size brought by different obfuscating options on datasets, where T and BS refer to executing time and binary size, respectively, and $O_1$ and $O_2$ stand for identical meaning in \S\ref{sec:evaluation:rq2:static}.}
    \label{table:overhead}
    \resizebox{0.5\columnwidth}{!}{%
        \begin{tabular}{@{}ccccccccc@{}}
            \toprule
            \textbf{}                                                                & \multicolumn{2}{c}{\textbf{D$_1$}} & \multicolumn{2}{c}{\textbf{D$_2$}}                    & \multicolumn{2}{c}{\textbf{D$_4$}} & \multicolumn{2}{c}{\textbf{D$_5$}}                                                                                                                                                  \\
                                                                                     & \textbf{T}                               & \begin{tabular}[c]{@{}c@{}}\textbf{BS}\end{tabular} & \textbf{T}                               & \begin{tabular}[c]{@{}c@{}}\textbf{BS}\end{tabular} & \textbf{T} & \begin{tabular}[c]{@{}c@{}}\textbf{BS}\end{tabular} & \textbf{T} & \begin{tabular}[c]{@{}c@{}}\textbf{BS}\end{tabular}
            \\ \midrule
            original                                                                 & 1                                  & 1                                                     & 1                                  & 1                                                     & 1    & 1                                                     & 1    & 1                                                     \\ \midrule
            \begin{tabular}[c]{@{}c@{}}name\\ obfuscation \end{tabular}                & 0.96                               & 1.00                                                  & 0.98                               & 0.96                                                  & 1.01 & 1.00                                                  & 0.99 & 1.02                                                  \\
            \begin{tabular}[c]{@{}c@{}}memory\\ obfuscation \end{tabular}              & 1.11                               & 1.22                                                  & 1.47                               & 1.04                                                  & 1.14 & 1.24                                                  & 0.99 & 1.00                                                  \\ \midrule
            flattening                                                               & 1.02                               & 1.09                                                  & 1.08                               & 1.01                                                  & 1.08 & 1.14                                                  & 1.01 & 1.17                                                  \\
            \begin{tabular}[c]{@{}c@{}}flattening\\ ($O_1$)\end{tabular}             & 1.06                               & 1.13                                                  & 1.19                               & 1.02                                                  & 1.18 & 1.14                                                  & 1.13 & 1.18                                                  \\
            \begin{tabular}[c]{@{}c@{}}flattening\\ ($O_2$)\end{tabular}             & 1.11                               & 1.14                                                  & 1.21                               & 1.03                                                  & 1.27 & 1.24                                                  & 1.29 & 1.36                                                  \\  \midrule
            alias disruption                                                   & 1.03                               & 1.06                                                  & 1.09                               & 1.00                                                  & 1.07 & 1.02                                                  & 1.00 & 1.02                                                  \\
            \begin{tabular}[c]{@{}c@{}}alias disruption\\ ($O_1$)\end{tabular} & 1.09                               & 1.25                                                  & 1.23                               & 1.08                                                  & 1.08 & 1.23                                                  & 1.15 & 1.42                                                  \\
            \begin{tabular}[c]{@{}c@{}}alias disruption\\ ($O_2$)\end{tabular} & 1.28                               & 1.26                                                  & 1.28                               & 1.08                                                  & 1.28 & 1.23                                                  & 1.19 & 1.43                                                  \\

            \bottomrule
        \end{tabular}%
    }
\end{table}

As we can see, the name obfuscation nearly brings no overhead in both runtime and binary size. This is because name obfuscation totally targets on the function names in the custom section. For a virtual machine, whether the function name is readable has no impact on executing speed. We believe that fluctuations in the numbers at the 4th row are the result of measurement errors.
As for the memory obfuscation, its imported overhead fluctuates significantly.
Interestingly, we can observe that it has no impact on cases in \textbf{D$_5$}. This is because they are constructed manually, for simplicity, authors do not introduce any interactions with the memory.
For the other three datasets, we can easily conclude that the overhead brought by the memory obfuscation has different impact on consumed time and binary size. For example, cases in \textbf{D$_2$} interacted with the environment by command-line arguments, indicating lots of memory manipulation operations. Therefore, around 47\% overhead is imported in terms of executing time. However, compared to its total number of instructions (tens of thousands instructions for each case), the overhead of binary size by hooking memory related instructions can be neglected.

Moreover, as for code obfuscation, even under the most aggressive option, it will only import less than 30\% overheads, except for the artificially constructed \textbf{D$_5$}.
From Table~\ref{table:effectiveness} in \S\ref{sec:evaluation:rq2:static}, we can conclude that the obfuscation effect of $O_1$ is enough for hindering the current state-of-the-art symbolic executors. Under this scenario, {\framework} can only introduce less than 20\% overhead in both consumed time and binary size, which we think is a balance point between security and performance.
Therefore, it is shown that \textit{{\framework} can introduce less than 30\% overhead in terms of binary size and executing time under the most aggressive obfuscating options. Considering the balance between efficiency and effectiveness, {\framework} can achieve a better performance}.

\begin{tcolorbox}[title= \textbf{RQ-3} Answer, left=2pt, right=2pt, top=2pt, bottom=2pt]
Name obfuscation brings no overhead, while memory obfuscation brings a higher overhead on command-line interacting Wasm binaries in terms of executing time. Moreover, considering the balance between effectiveness and efficiency, code obfuscation under $O_1$ brings around no more than 20\% overhead in both terms of consumed time and binary size.
\end{tcolorbox}

\section{Related Work}

\noindent \textbf{WebAssembly Binary.}
There are many works based on the Wasm binaries, covering various aspects, e.g., program analysis~\cite{wasabi,debug-info,slice,wasmati,eosafe}, malware detection and evasion~\cite{SEISMIC,minerray,binary-mutate1}, and testing~\cite{wafl,fuzzm}.
Specifically, Wasabi~\cite{wasabi} is a dynamic analysis framework for Wasm which is able to perform instructions instrumentation on Wasm binaries. To obtain analysis results, Wasabi also inserts some helper functions as imported ones, which take the instrumented instructions as input.
Moreover, to detect malicious cryptomining programs, SEISMIC~\cite{SEISMIC} monitors the frequency of specific instructions in the given Wasm binary at runtime by instrumenting them.
In addition, J.C. Arteaga et al.~\cite{binary-mutate1} implements Wasm-mutate to perform malware evasion on Wasm binaries. It provides many predefined strategies, e.g., instructions replacement and module structure transformation.

\noindent \textbf{Code Obfuscation.}
Obfuscation has been widely adopted for decades to resist human reverse engineering and program analysis techniques. Lots of representative work were proposed before~\cite{flatten,obfuscation,data-obfuscation,tigress,proteus,ollvm,collatz,against-symbol,svp,low-cost,loki,opaque-predicate}.
For example, Proteus~\cite{proteus} adopts the virtual machine technology to perform obfuscation by translating the original program to a new instruction set used by the virtual machine.
Tigress~\cite{tigress} is a source-to-source obfuscator for C. It supports many traditional obfuscation methods, e.g., control flow obfuscation, and virtualization.
Due to the development of program analysis techniques, countering symbolic execution should be taken into consideration when implementing an obfuscator~\cite{collatz,against-symbol,svp,loki}.
Specifically, S. Banescu et al.~\cite{against-symbol} apply range dividers and input variants to deliberately increase the number of feasible paths, and M. Ollivier et al.~\cite{svp} thoroughly study path-oriented protections to hinder symbolically executing.
In addition, there are many efforts aimed at SMT solvers used in symbolic execution~\cite{mba,loki,low-cost}. For example, M. Schloegel et al.~\cite{loki} proposed a general framework for synthesizing MBA expressions that can effectively address current anti-obfuscation attacks.
As for obfuscation on Wasm, there is only one work~\cite{first} which evaluated if source-level obfuscation can still evade malware detection on compiled Wasm binaries. However, they do not propose any obfuscator and our results mentioned in \S\ref{sec:rq2:malware} are better than theirs.

\section{Threats of Validity}

\noindent \textbf{Deobfuscation.}
Pattern attack is a method that recognizes specific code structures generated by obfuscations and deobfuscates them. To avoid being deobfuscated, a common method is to increase the code diversity, which can be easily achieved on binary level.
For example, in control flow flattening, we can insert dead code blocks in each layer to diversify the code structure.
Moreover, to avoid the opaque predicates being identified, we can construct multiple variants based on our proposed Collatz-based one (like introducing hash function), and insert one of them before \texttt{call\_indirect}.

Against Wasm, Binaryen~\cite{binaryen} is the only usable tool to conduct deobfuscation to some extent.
Specifically, it will lift the Wasm bytecode to its custom IR, perform optimizations, and compile it back to Wasm.
Not surprisingly, according to our experimental results, Binaryen can deobfuscate around 70\% obfuscation effect of control flow flattening and alias disruption by counting the number of flattened code blocks and \texttt{call\_indirect} instructions.
However, such a deobfuscation turns totally invalid when introducing Collatz-based opaque predicates to the above obfuscation methods.
Such a situation is evaluated in \S\ref{sec:rq3} under the $O_1$ option.
As we can see, the overhead in both terms of executing time and binary size increased only 14\% and 18\%, respectively, which is acceptable for a more robust obfuscation.

\noindent \textbf{Potential obfuscation method.}
In addition to five proposed obfuscating methods, we emphasize that other obfuscating methods can be easily achieved through extending {\framework}.
For example, the code block splitting and code block rearranging method mentioned in \S\ref{sec:approach:code-obfuscation:flattening} can serve as the prerequisite for other control flow obfuscations, such as random control flow and bogus control flow.
Moreover, rearranging these split code blocks can also achieve other obfuscating goals.
Additionally, the \textit{Mixed Boolean-Arithmetic} (MBA)~\cite{mba} approach that is often used to resist symbolic execution can also be easily introduced in Wasm due to the stack-based format of Wasm. Once an MBA expression is converted to the AST format, it can be easily inserted into Wasm binaries.

\noindent \textbf{Benchmark.}
When evaluating the effectiveness of {\framework}, we select different datasets for different jobs, which is mainly due to the limited functionalities of these tools.
For example, although manticore can symbolically execute Wasm binaries, it does not support WASI (used by \textbf{D$_1$} and \textbf{D$_2$}) and customized import functions (\textbf{D$_3$}). Similarly, wasp can only support a customized set of instructions, i.e., \textbf{D$_5$}.
Moreover, we claim that these datasets are representative.
Specifically, they cover both standalone (\textbf{D$_1$} and \textbf{D$_2$}) and web-integrated format (\textbf{D$_3$}). They are either compiled from source code (\textbf{D$_1$} to \textbf{D$_4$}) or even composed manually (\textbf{D$_5$}). Moreover, there are simple algorithms ranging from sequence searching to string operations (\textbf{D$_1$}), and complicated real-world applications (\textbf{D$_2$}).
Therefore, we believe that these five datasets are representative enough, and are not designed specifically for evaluating tasks.

\section{Conclusion}
This paper presents the first general-purpose Wasm binary obfuscation framework, which works on both data-level and code-level obfuscation. 
Results show that {\framework} can not only keep the original semantics intact, but also be effective in resisting human reverse engineering, hiding original intents, and hindering the analysis from state-of-the-art static analyzers. Moreover, it brings in acceptable overhead in terms of both executing time and binary size. Our research has shed light on the promising direction of Wasm binary research, including Wasm code protection, Wasm binary diversification, and the attack-defense arm race of Wasm binaries.

\bibliographystyle{ACM-Reference-Format}
\bibliography{chaos-reference}


\end{document}